\documentclass[equations,10pt]{article}
\usepackage{amssymb,amsmath,amsthm}
\usepackage{eucal}
\usepackage{color}

\renewcommand{\theequation}{\arabic{equation}}
\begin{document}
\bibliographystyle{plain}
\def\m@th{\mathsurround=0pt}
\mathchardef\bracell="0365
\def\upbrall{$\m@th\bracell$}
\def\undertilde#1{\mathop{\vtop{\ialign{##\crcr
    $\hfil\displaystyle{#1}\hfil$\crcr
     \noalign
     {\kern1.5pt\nointerlineskip}
     \upbrall\crcr\noalign{\kern1pt
   }}}}\limits}
\def\theequation{\arabic{section}.\arabic{equation}}
\newcommand{\ar}{\alpha}
\newcommand{\aar}{\bar{a}}
\newcommand{\bb}{\beta}
\newcommand{\gm}{\gamma}
\newcommand{\Gm}{\Gamma}
\newcommand{\en}{\epsilon}
\newcommand{\ven}{\varepsilon}
\newcommand{\dd}{\delta}
\newcommand{\sg}{\sigma}
\newcommand{\kp}{\kappa}
\newcommand{\ld}{\lambda}
\newcommand{\oa}{\omega}
\newcommand{\be}{\begin{equation}}
\newcommand{\ee}{\end{equation}}
\newcommand{\bea}{\begin{eqnarray}}
\newcommand{\eea}{\end{eqnarray}}
\newcommand{\bse}{\begin{subequations}}
\newcommand{\ese}{\end{subequations}}
\newcommand{\nn}{\nonumber}
\newcommand{\vf}{\varphi}
\newcommand{\sn}{{\rm sn}}
\newcommand{\wh}{\widehat}
\newcommand{\ol}{\overline}
\newcommand{\wt}{\widetilde}
\newcommand{\ut}{\undertilde}
\newcommand{\ip}{{i^\prime}}
\newcommand{\jp}{{j^\prime}}
\newcommand{\cn}{{\rm cn}}
\newcommand{\dn}{{\rm dn}}
\newcommand{\btP}{\,^{t\!}\boldsymbol{P}}
\newcommand{\tLd}{\,^{t\!}\boldsymbol{\Lambda}}
\newcommand{\ca}{{\frak a}}
\newcommand{\ck}{{\frak k}}
\newcommand{\cp}{{\frak p}}
\newcommand{\cA}{{\frak A}}
\newcommand{\cK}{{\frak K}}
\newcommand{\bI}{\boldsymbol{I}}
\newcommand{\bO}{\boldsymbol{O}}
\newcommand{\bA}{\boldsymbol{A}}
\newcommand{\bB}{\boldsymbol{B}}
\newcommand{\bV}{\boldsymbol{V}}
\newcommand{\bZ}{\boldsymbol{Z}}
\newcommand{\bU}{\boldsymbol{U}}
\newcommand{\bC}{\boldsymbol{C}}
\newcommand{\bOm}{\pmb{\Omega}}
\newcommand{\bKp}{\boldsymbol{\Kappa}}
\newcommand{\buk}{\boldsymbol{u}_k}
\newcommand{\bul}{\boldsymbol{u}_\ell}
\newcommand{\tII}{\,^{t\!}\boldsymbol{I}}
\newcommand{\tuk}{\,^{t\!}\boldsymbol{u}_{k^\prime}}
\newcommand{\tul}{\,^{t\!}\boldsymbol{u}_{\ell^\prime}}
\newcommand{\tull}{\,^{t\!}{\bf u}_{-\ell+\ld}}
\newcommand{\tuq}{\,^{t\!}{\bf u}_{-q_j+\ld}}
\newcommand{\tcl}{\,^{t\!}\boldsymbol{c}_{\ell}}
\newcommand{\tck}{\,^{t\!}\boldsymbol{c}_{k^\prime}}
\newcommand{\ssk}{\sigma_{k^\prime}}
\newcommand{\ssl}{\sigma_{\ell^\prime}}
\newcommand{\pte}{(\partial_t-\partial_\eta)}
\newcommand{\pxe}{(\partial_x-\partial_\eta)}
\newcommand{\dint}{\int_\Gamma d\mu(\ell) }
\def\hypotilde#1#2{\vrule depth #1 pt width 0pt{\smash{{\mathop{#2}
\limits_{\displaystyle\widetilde{}}}}}}
\def\hypohat#1#2{\vrule depth #1 pt width 0pt{\smash{{\mathop{#2}
\limits_{\displaystyle\widehat{}}}}}}
\def\hypo#1#2{\vrule depth #1 pt width 0pt{\smash{{\mathop{#2}
\limits_{\displaystyle{}}}}}}

\newcommand{\pii}{P$_{{\rm\small II}}$}  
\newcommand{\pvi}{P$_{{\rm\small VI}}$}   
\newcommand{\ptf}{P$_{{\rm\small XXXIV}}$}   
\newcommand{\diffE}{O$\triangle$E}   

\newcommand{\pp}{\partial}
\newcommand{\hf}{\frac{1}{2}}
\newcommand{\bR}{\boldsymbol{R}}
\newcommand{\bS}{\boldsymbol{S}}
\newcommand{\ith}{$i^{\rm th}$\ }
\newcommand{\bu}{{\boldsymbol u}}
\newcommand{\bell}{{\boldsymbol l}}
\newcommand{\bbS}{{\boldsymbol S}}
\newcommand{\bj}{{\boldsymbol j}}
\newcommand{\bt}{{\boldsymbol t}}
\newcommand{\bm}{{\boldsymbol m}}
\newcommand{\boa}{{\boldsymbol \omega}}
\newcommand{\bW}{\bar{W}}
 \newcommand{\pl}{\partial}
 \newcommand{\ddp}{\frac{\partial}{\partial p}}
 \newcommand{\ddq}{\frac{\partial}{\partial q}}
 \newcommand{\ddr}{\frac{\partial}{\partial r}}
 \newcommand{\Ld}{\pmb{\Lambda}}
 \newcommand{\I}{\boldsymbol{I}}
 \newcommand{\bP}{\boldsymbol{P}}
 \newcommand{\tbP}{\,^{t\!}\boldsymbol{P}}
 \newcommand{\tbC}{\,^{t\!}{\bf C}}
 \newcommand{\ddint}{\int_\Gamma d\ld(\ell) }
 \newcommand{\vE}{\vec{E} }
 \newcommand{\vL}{\vec{L} }
 \newcommand{\vn}{\vec{n} }
 \newcommand{\vR}{\vec{R} }
 \newcommand{\vP}{\vec{P} }
 \newcommand{\vna}{\vec{\nabla} }
 \newcommand{\vv}{\vec{v} }
 \newcommand{\vF}{\vec{F} }
 \newcommand{\vj}{\vec{j} }
 \newcommand{\vB}{\vec{B} }
 \newcommand{\vr}{\vec{r} }
 \newcommand{\vp}{\vec{p} }
 \newcommand{\vk}{\vec{k} }
\newcommand{\mbe}{{\boldsymbol e}}
\newcommand{\bE}{{\boldsymbol E}}
\newcommand{\bnab}{{\boldsymbol \nabla}}
\newcommand{\buu}{{\boldsymbol u}}
\newcommand{\bv}{{\boldsymbol v}}
\newcommand{\ba}{{\boldsymbol a}}
\newcommand{\bbb}{{\boldsymbol b}}
\newcommand{\bJ}{{\boldsymbol J}}
\newcommand{\bc}{{\boldsymbol c}}
\newcommand{\bw}{{\boldsymbol w}}
\newcommand{\mbx}{{\boldsymbol x}}
\newcommand{\mby}{{\boldsymbol y}}
\newcommand{\bz}{{\boldsymbol z}}
\newcommand{\brr}{{\boldsymbol r}}
\newcommand{\bp}{{\boldsymbol p}}
\newcommand{\bk}{{\boldsymbol k}}
\newcommand{\btt}{{\boldsymbol t}}
\newcommand{\bmm}{{\boldsymbol m}}
\newcommand{\bdd}{{\boldsymbol \delta}}
\newcommand{\bze}{{\boldsymbol 0}}
\newcommand{\boma}{{\boldsymbol \omega}}
\newcommand{\bet}{{\boldsymbol \eta}}
\newcommand{\bxi}{{\boldsymbol \xi}}
 \newcommand{\mbv}{\boldmath{v}}
 \newcommand{\mbxi}{\boldmath{\xi}}
 \newcommand{\mbeta}{\boldmath{\eta}}
 \newcommand{\mbw}{\boldmath{w}}
 \newcommand{\mbu}{\boldmath{u}}

 \def\hypotilde#1#2{\vrule depth #1 pt width 0pt{\smash{{\mathop{#2}
 \limits_{\displaystyle\widetilde{}}}}}}
 \def\hypohat#1#2{\vrule depth #1 pt width 0pt{\smash{{\mathop{#2}
 \limits_{\displaystyle\widehat{}}}}}}
 \def\hypo#1#2{\vrule depth #1 pt width 0pt{\smash{{\mathop{#2}
 \limits_{\displaystyle{}}}}}}

\newtheorem{theorem}{Theorem}[section]
\newtheorem{lemma}{Lemma}[section]
\newtheorem{cor}{Corollary}[section]
\newtheorem{prop}{Proposition}[section]
\newtheorem{definition}{Definition}[section]
\newtheorem{conj}{Conjecture}[section]

\begin{flushright}
\today \\
\end{flushright}
\begin{center}
{\large{\bf Semi-classical Laguerre polynomials and a third order discrete integrable equation}}
\vspace{.4cm}

Paul E. Spicer \\
{\it Katholieke Universiteit Leuven, Departement Wiskunde, Celestijnenlaan 200B
B-3001 Leuven, Belgium} \\
Email: Paul.Spicer@wis.kuleuven.be \\
\vspace{.4cm}
and \\
\vspace{.4cm}
Frank W. Nijhoff\\
{\it Department of Applied Mathematics, University of Leeds, Leeds LS2 9JT, U.K.} \\
Email: frank.nijhoff@gmail.com \\
 \vspace{.2cm}

\end{center}

\vspace{.4cm}
\centerline{\bf Abstract}
\vspace{.2cm}

\noindent

A semi-discrete Lax pair formed from the differential system and recurrence relation for semi-classical orthogonal polynomials, leads
to a discrete integrable equation for a specific semi-classical orthogonal polynomial weight.  The main example we use is a semi-classical
Laguerre weight to derive a third order difference equation with a corresponding Lax pair.

\section{Introduction}
\setcounter{equation}{0}

The connection between semi-classical orthogonal polynomials and discrete integrable systems is well established.  The earliest
example of a discrete integrable system in semi-classical orthogonal polynomials can be attributed first to Shohat in 1939 \cite{Shohat},
then second by Freud \cite{Freud} in 1976. However it wasn't until the 1990's, when the focus within integrable systems shifted from
continuous to discrete, that Fokas, Its, Kitaev, \cite{Fokas} gave this equation a name; discrete Painlev\'e I,  (d-P$_{{\rm\small
I}}$).

Since then, other examples of discrete Painlev\'e equations have been found through exploring the recursive structures of
different semi-classical orthogonal polynomial families, including semi-classical Hermite \cite{MagnusF}, semi-classical Laguerre \cite{FW3} and
semi-classical Charlier \cite{van Assche}.

We define an orthogonal polynomial sequence $\{P_{n}(z)\}_{n=0}^{\infty}$ with respect to a weight function $w(z)$
on an interval $(a,b)$ as
\be
\int_{a}^{b}P_{n}(z)P_{m}(z)w(z)dz=h_{n}\dd_{nm}\ ,
\ee
with the corresponding recurrence relation
\be
zP_{n}(z)=P_{n+1}+S_{n}P_{n}+R_{n}P_{n-1}\label{eq:Rec}\
\ee
for a monic orthogonal polynomial family $P_{n}(z)=z^{n}+p_{nn-1}z^{n-1}+p_{nn-2}z^{n-2}+\ldots$
From Bochner \cite{Pearson} we know that if $\{P_{n}(z)\}$ is a
sequence of {\it classical} orthogonal polynomials, then $P_{n}(x)$
is a solution of the second-order differential equation \be
\phi(z)\frac{d^{2}y}{dz^{2}}+\psi(z)\frac{dy}{dz}=\lambda_{n}y \ee
where  $\phi(z)$ and $\psi(z)$ are fixed polynomials of degree $\leq
2$ and $\leq 1$ respectively, and $\lambda_{n}$ is a real number
depending on the degree of the polynomial solution. As a consequence
of this the weights of classical orthogonal polynomials satisfy a
first order differential equation called the Pearson differential
equation \be\label{eq:Pear} \frac{d}{dz}(\phi(z) w(z))= \psi(z)
w(z)\ , \ee when the degrees of $\phi$ and $\psi$ satisfy deg$\
\phi\leq 2$ and deg$\ \psi=1$ . However when the deg$\ \phi > 2$
and$\backslash$or deg$\ \psi >1$ then the weight function produces a
class of semi-classical orthogonal polynomials.

Our approach to semi-classical orthogonal polynomials is to make use of the Laguerre method \cite{Laguerre}
({\it not} to be confused with Laguerre orthogonal polynomials),
which derives a pair of first order differential equations for a general class of orthogonal polynomials, after the reduction of continued fractions.
The connection with semi-classical orthogonal polynomials occurs because we associate the system with
a semi-classical weight function $w(x)$ of the polynomials with the Pearson equation (\ref{eq:Pear}).
For convenience we choose to write the Pearson equation in the following form
\begin{equation}
W(z)\pp_{z}w(z)=V(z)w(z)\label{eq:VW}\ ,
\end{equation}
where $V(z)=\psi-\phi'$ and $W(z)=\phi$.
While our aim and approach is different, the Laguerre method  has been used to find
connections with integrable systems, including continuous Painlev\'e equations, recently.
Magnus \cite{Magnus}, found a continuous Painlev\'e equation of the sixth kind from the recurrence
coefficients of a semi-classical Jacobi polynomial and Forrester and Witte \cite{FW1,FW2}, found
a Painlev\'e equation of the fifth kind, also using the Laguerre method, but one that has been extended to
include bi-orthogonal polynomials.

Our work will consider a semi-classical Laguerre weight, similar to that used by \cite{FW3}.  The semi-classical Laguerre polynomials have not been
as widely explored as the semi-classical Hermite polynomials, nor are they as complex as the semi-classical Jacobi polynomials.  Thus,
Laguerre polynomials are an appropriate choice for finding new discrete integrable systems.


In section 2 we use the so-called Laguerre method to derive the differential system for semi-classical (monic) orthogonal
polynomials.  We show how the compatibility between the differential system and the recurrence relation (\ref{eq:Rec})
leads to a semi-discrete Lax equation \cite{Lax}, from which discrete integrable systems can be derived for specific semi-classical orthogonal polynomial weights.
In section 3 we choose the semi-classical Laguerre weight $l_{0}(x)=(x-t)^{\ar}e^{-(ax+\frac{b}{2}x^{2})}$ which leads to a coupled difference system and
a corresponding third order nonlinear difference equation.

\section{The Laguerre Method}
\setcounter{equation}{0}



We introduce a moment generating function, the Stieltjes function,
\begin{equation}
f(z) = \int \frac{w(x)}{z-x}d(x)\label{eq:fz}
\end{equation}
(Stieltjes transform of the orthogonality measure $w(x)$) then equations for $P_{n}$ can be summarized as
\begin{equation}
f(z)P_{n}(z)=P_{n-1}^{(1)}(z)+\en_{n}(z)\label{eq:fp}  ,
\end{equation}
where $P_{n-1}^{(1)}(z)$ is an associated polynomial to $P_{n}(z)$,
with degree $n-1$.  Although $\en_{n}(z)$ is not a polynomial, we
can define it as
\begin{equation}
  \en_{n}(z)  = \int \frac{P_{n}(x)}{z-x}w(x)dx  \label{eq:enz}.
\end{equation}
The polynomials $P_{n-1}^{(1)}(z)$, as well as the $\en_{n}(z)$,
satisfy the same recurrence relations (\ref{eq:Rec}), but with
$P_{-1}^{(1)}(z)=0$. Additionally we have the following relations
between $P_{n}, P_{n}^{(1)}$ and $\en_{n}$ \bse\begin{eqnarray}
P_{n}P_{n-2}^{(1)}-P_{n-1}P_{n-1}^{(1)} & = & -h_{n-1} \label{eq:hnP}\\
P_{n-1}\en_{n}-P_{n}\en_{n-1} & = & -h_{n-1} \label{eq:hnE}\ ,
\end{eqnarray}\ese
which can be found using the Christoffel-Darboux identity.
Since both $P_{n}(z)$ and $\en_{n}(z)$ satisfy the recurrence relation (\ref{eq:Rec}) we can give
an explicit form of $P_{n}(z)$ and $\en_{n}(z)$ defined in terms of the recurrence relation's coefficients:
\bse\begin{eqnarray}
P_{n}(z)&=&z^{n}-\left(\sum_{j=0}^{n-1}S_{j}\right)z^{n-1}+\sum_{j=1}^{n-1}\left(\sum_{k=0}^{j-1}S_{j}S_{k}-R_{j}\right)z^{n-2}+\cdots \label{eq:edPn} \\
\en_{n}(z)&=&h_{n}\left(\frac{1}{z^{n+1}}+\left(\sum_{j=0}^{n}S_{j}\right)\frac{1}{z^{n+2}}+\sum_{j=0}^{n}\left(R_{j+1}+\sum_{i=0}^{j}S_{j}S_{i}\right)\frac{1}{z^{n+3}}+\cdots\right)\label{eq:eden}\
.
\end{eqnarray}\ese

Semi-classical orthogonal polynomials may be defined through a differential difference equation of the form
\begin{equation}
W(z)\pp_{z}f(z)=V(z)f(z)+U(z)\label{eq:lde}
\end{equation}
which comes from considering $W(z)(\pp_{z}f(z))$ and the Pearson equation (\ref{eq:VW}).
\begin{eqnarray*}
W(z)(\pp_{z}f(z)) = -\int \frac{W(z)w(x)}{(z-x)^{2}}dx & = & -\int\frac{d}{dx}\left(\frac{1}{z-x}W(z)w(x)\right)dx + \int \frac{W(z)}{z-x}\pp_{x}w(x)  \\
                  & = & \int\frac{W(z)}{W(x)}V(x)\frac{1}{z-x}w(x) dx  \\
                  & = & V(z)f(z)+W(z)\int\left(\frac{V(x)}{W(x)}-\frac{V(z)}{W(z)}\right)\frac{w(x)}{z-x} dx\
\end{eqnarray*}
On the first line we assume the first term reduces to zero because of parameter constraints and then we have that:
\begin{equation*}
   U(z)  =  W(z)\int\left(\frac{V(x)}{W(x)}-\frac{V(z)}{W(z)}\right)\frac{w(x)}{z-x} dx\ ,
\end{equation*}
where $U(z)$ is a polynomial in $z$.

\subsection{The fundamental linear system for semi-classical orthogonal polynomials}

We start with the equation (\ref{eq:fp}), differentiate it and multiply by $W$, so that we can then
make use of the first order linear differential equation (\ref{eq:lde})
(with the exception, that for this case we will consider the $x$ variable to be dominant).
\begin{eqnarray}
                                                    Wf\pp_{x}P_{n}+(Vf+U)P_{n} & = & W(\pp_{x}P^{(1)}_{n-1}+\pp_{x}\en_{n}) \nn \\
W\pp_{x}P_{n}(P^{(1)}_{n-1}+\en_{n})+VP_{n}(P^{(1)}_{n-1}+\en_{n})+UP_{n}^{2}  & = & W(\pp_{x}P^{(1)}_{n-1}+\pp_{x}\en_{n})P_{n}\label{eq:TPE}
\end{eqnarray}
We then go about separating the polynomial expression $P_{n-1}^{(1)}$ and $\en_{n}$ so we get the following two equivalent expressions,
which we denote $\Theta_{n}$
\bse\begin{eqnarray}
\Theta_{n} & = & W(\pp_{x}P_{n-1}^{(1)}P_{n}-\pp_{x}P_{n}P_{n-1}^{(1)})-UP_{n}^{2}-VP_{n}P_{n-1}^{(1)}\label{eq:TP} , \\
           & = & W(\pp_{x}P_{n}\en_{n}-\pp_{x}\en_{n}P_{n})+VP_{n}\en_{n}\label{eq:TE},
\end{eqnarray}\ese
where $\Theta_{n}$ is a polynomial bounded by a constant.
We try the same method again except this time we use $fP_{n-1}$, which is again differentiated and multiplied by $W$.
\begin{eqnarray}
                                                     \pp_{x}fP_{n-1}+f\pp_{x}P_{n-1} & = & \pp_{x}P_{n-2}^{(1)}+\pp_{x}\en_{n-1}   \\
VP_{n-1}(P_{n-1}^{(1)}+\en_{n})+UP_{n}P_{n-1}+W\pp_{x}P_{n-1}(P_{n-1}^{(1)}+\en_{n}) & = & W(\pp_{x}P_{n-2}^{(1)}+\pp_{x}\en_{n-1})P_{n} \nn \\
                                                                                     &   & \label{eq:Om}
\end{eqnarray}
Again we separate the polynomial expression $P_{n-1}^{(1)}$ and $\en_{n}$ to get a second object, which will be called $\Omega_{n}$:
\bse\begin{eqnarray}
\Omega_{n} & = & W(P_{n}\pp_{x}P_{n-2}^{(1)}-P_{n-1}^{(1)}\pp_{x}P_{n-1})-VP_{n-1}P_{n-1}^{(1)}-UP_{n}P_{n-1}\label{eq:OP}  \\
           & = & W(\en_{n}\pp_{x}P_{n-1}-P_{n}\pp_{x}\en_{n-1})+V\en_{n}P_{n-1}\label{eq:OE}
\end{eqnarray}\ese
We can express both $\Omega_{n}$ and $\Theta_{n}$ in terms of the recurrence coefficients by substituting the expressions for $P_{n}$
(\ref{eq:edPn}) and $\en_{n}$ (\ref{eq:eden}) into $\Omega_{n}$ (\ref{eq:OE}) and $\Theta_{n}$ (\ref{eq:TE}).
{\small\bse\begin{eqnarray}\label{eq:Theta}
\Theta_{n} & = & W(x)h_{n}\left\{\left[\frac{1}{x^{n+1}}+\left(\sum_{j=0}^{n}S_{j}\right)\frac{1}{x^{n+2}}+\cdots\right]\times\left[nx^{n-1}-\left(\sum_{j=0}^{n-1}S_{j}\right)(n-1)x^{n-2}+\cdots\right]\right. \nn \\
           &   & +\left.\left[\frac{n+1}{x^{n+2}}+\left(\sum_{j=0}^{n}S_{j}\right)\frac{n+2}{x^{n+3}}+\cdots\right]\times\left[x^{n}-\left(\sum_{j=0}^{n-1}S_{j}\right)x^{n-1}+\cdots\right]\right\}\nn \\
           &   & +V(x)\times h_{n}\left[\frac{1}{x^{n+1}}+\left(\sum_{j=0}^{n}S_{j}\right)\frac{1}{x^{n+2}}+\cdots\right]\times \left[x^{n}-\left(\sum_{j=0}^{n-1}S_{j}\right)x^{n-1}+\cdots\right]\nn  \\
           &   &
\end{eqnarray}
\begin{eqnarray}\label{eq:Omega}
\Omega_{n} & = & W(x)\left\{h_{n}\left[\frac{1}{x^{n+1}}+\left(\sum_{j=0}^{n}S_{j}\right)\frac{1}{x^{n+2}}+\sum_{j=0}^{n}\left(R_{j+1}+\sum_{k=0}^{j}S_{j}S_{k}\right)\frac{1}{x^{n+3}}+\cdots\right]\right.\nn  \\
           &   & \left.\times\left[(n-1)x^{n-2}-(n-2)\left(\sum_{j=0}^{n-2}S_{j}\right)x^{n-3}+(n-3)\sum_{j=1}^{n-2}\left(\sum_{k=0}^{j-1}S_{j}S_{k}-R_{j}\right)x^{n-4}+\cdots\right]\right. \nn \\
           &   & +\left.h_{n-1}\left[x^{n}-\left(\sum_{j=0}^{n-1}S_{j}\right)x^{n-1}+\sum_{j=1}^{n-1}\left(\sum_{k=0}^{j-1}S_{j}S_{k}-R_{j}\right)x^{n-2}+\cdots\right]\right. \nn \\
           &   & \left.\times\left[\frac{n}{x^{n+1}}+\left(\sum_{j=0}^{n-1}S_{j}\right)\frac{(n+1)}{x^{n+2}}+\sum_{j=0}^{n-1}\left(R_{j+1}+\sum_{k=0}^{j}S_{j}S_{k}\right)\frac{(n+2)}{x^{n+3}}+\cdots\right]\right\} \nn \\
           &   & +V(x)  \nn\\
           &   & \times h_{n}\left[\frac{1}{x^{n+1}}+\left(\sum_{j=0}^{n}S_{j}\right)\frac{1}{x^{n+2}}+\sum_{j=0}^{n}\left(R_{j+1}+\sum_{k=0}^{j}S_{j}S_{k}\right)\frac{1}{x^{n+3}}+\cdots\right]\nn  \\
           &   & \times\left[x^{n-1}-\left(\sum_{j=0}^{n-2}S_{j}\right)x^{n-2}+\sum_{j=1}^{n-2}\left(\sum_{k=0}^{j-1}S_{j}S_{k}-R_{j}\right)x^{n-3}+\cdots\right] .
\end{eqnarray}\ese}
These definitions will be particularly useful when we are looking at examples of specific semi-classical weights.

Since the recurrence relation (\ref{eq:Rec}) can be expressed in a matrix form
\begin{equation}
\psi_{n+1}(x) = \left(\begin{array}{cc}
x-S_{n} & -R_{n}  \\
      1 & 0
\end{array}\right)\psi_{n}(x)
,\  {  \rm where \ } \psi_{n}(x)=\left(\begin{array}{c}
P_{n}(x) \\
P_{n-1}(x)
\end{array}\right)\label{eq:RecN}
\end{equation}
we collect the important relations we have derived so far and put them in a matrix form so that our
intended differential system can be written as one expression.
We begin with the two expressions (\ref{eq:TP}) and (\ref{eq:OP}),
written in matrix form:
\bse\begin{equation}
\left(\begin{array}{cc}
P_{n-1} & -P_{n-2}^{(1)}  \\
P_{n}   & -P_{n-1}^{(1)}
\end{array}\right)
\left(\begin{array}{c}
W\pp_{x}P^{(1)}_{n-1} \\
W\pp_{x}P_{n}
\end{array}\right) =
\left(\begin{array}{c}
\Omega_{n}+VP_{n-1}P_{n-1}^{(1)}+UP_{n}P_{n-1} \\
\Theta_{n}+VP_{n}P_{n-1}^{(1)}+UP_{n}^{2}
\end{array}\right),
\end{equation}
which can easily be solved making use of (\ref{eq:hnP}) to give:
\begin{equation}
\left(\begin{array}{c}
W\pp_{x}P^{(1)}_{n-1} \\
W\pp_{x}P_{n}
\end{array}\right) = \frac{1}{h_{n-1}}
\left(\begin{array}{cc}
P_{n-1}^{(1)} & -P_{n-2}^{(1)}  \\
P_{n}         & -P_{n-1}
\end{array}\right)
\left(\begin{array}{c}
\Omega_{n}+VP_{n-1}P_{n-1}^{(1)}+UP_{n}P_{n-1} \\
\Theta_{n}+VP_{n}P_{n-1}^{(1)}+UP_{n}^{2}
\end{array}\right),
\end{equation}\ese
so that we have two differential equations:
\bse\begin{eqnarray}
W\pp_{x}P_{n}&=&\frac{1}{h_{n-1}}(\Omega_{n}P_{n}-\Theta_{n}P_{n-1})\label{eq:FDR} , \\
W\pp_{x}P_{n-1}^{(1)}&=&(\Omega_{n}P_{n-1}^{(1)}-\Theta_{n}P_{n-2}^{(1)}+Vh_{n-1}P_{n-1}^{(1)}+Uh_{n-1}P_{n}).
\end{eqnarray}\ese
Looking for a second differential relation for $P_{n}$, we take (\ref{eq:FDR}) with a reduced index in conjunction with the
recurrence relation (\ref{eq:Rec}), which leads to
\begin{equation}
W(\pp_{x}P_{n-1})=\frac{1}{h_{n-2}}\left(\Omega_{n-1}P_{n-1}-\frac{\Theta_{n-1}}{R_{n-1}}((x-S_{n-1})P_{n-1}-P_{n})\right)\label{eq:HSDR}.
\end{equation}
However we have no expression to remove the $x$ from the equation, so we consider the problematic part of the expression:
$(x-S_{n})\Theta_{n} = (x-S_{n})\left(W(\en_{n}\pp_{x}(P_{n})-\pp_{x}(\en_{n})P_{n})+V\en_{n}P_{n}\right)$,
which we expand using (\ref{eq:Rec}) and the differential of (\ref{eq:hnE}) to get:
\begin{eqnarray}
(x-S_{n})\Theta_{n} & = & W(-\pp_{x}\en_{n}(P_{n+1}+R_{n}P_{n-1})+\pp_{x}P_{n}(\en_{n+1}+R_{n}\en_{n-1}))\nn \\
                    &   & +VP_{n}(\en_{n+1}+R_{n}\en_{n-1}) \nn \\
                    & = & \Omega_{n+1}+R_{n}\Omega_{n}+Vh_{n}
\end{eqnarray}
This allows us to remove $x$ from (\ref{eq:HSDR}) to give a second differential equation.
\begin{equation}
W\pp_{x}P_{n-1}=\frac{1}{h_{n-1}}(\Theta_{n-1}P_{n}-\Omega_{n}P_{n-1})-VP_{n-1}\label{eq:SDR}
\end{equation}
We now have a {\it differential system}
\begin{equation}
W\pp_{x}\psi(x) = \frac{1}{h_{n-1}}
\left(\begin{array}{cc}
\Omega_{n}(x) & -\Theta_{n}(x)  \\
\Theta_{n-1}(x) & -(\Omega_{n}(x)+V(x)h_{n-1})
\end{array}\right)\psi(x)\label{eq:Diff}
\end{equation}
where $\psi(x)=\left(\begin{array}{c}
P_{n}(x) \\
P_{n-1}(x)
\end{array}\right)$.
Thus if we give the recurrence and differential equations in a semi-discrete Lax representation we have
\bse\label{eq:LP}\bea
\psi_{n+1}(x)&=&L_{n}(x)\psi_{n}(x)\\
\pp_{x}\psi_{n}(x) & = & M_{n}(x)\psi_{n}(x)
\eea\ese
where
\be
L_{n}=\left(\begin{array}{cc}
x-S_{n} & -R_{n}  \\
      1 & 0
\end{array}\right)\quad , \quad
M_{n}= \frac{1}{Wh_{n-1}}
\left(\begin{array}{cc}
\Omega_{n}(x) & -\Theta_{n}(x)  \\
\Theta_{n-1}(x) & -(\Omega_{n}(x)+V(x)h_{n-1})
\end{array}\right) \nn\ .
\ee
Here we have identified the Lax matrices $L_{n}$ and $M_{n}$.  So given a particular semi-classical weight we can identify the polynomials $V$ and $W$,
which in turn lead to expressions for $\Theta$ and $\Omega$.

\subsection{Compatibility relations}

We now use the differential system (\ref{eq:Diff}) with the matrix
form of the recurrence relation (\ref{eq:RecN}) in order to create a compatibility relation so that
relations between $\Omega_{n}$ and $\Theta_{n}$ can be derived.
Thus we consider the compatibility between the semi-discrete Lax pair, which leads to the semi-discrete Lax equation
\be\label{eq:Lax}
\pp_{x}L_{n}= M_{n+1}L_{n}-L_{n}M_{n}.
\ee
Equating this expression
\begin{eqnarray}
\left(\begin{array}{cc}
  1 & 0  \\
  0 & 0
\end{array}\right) & = &  \frac{1}{Wh_{n}} \left(\begin{array}{cc}
\Omega_{n+1}(x) & -\Theta_{n+1}(x)  \\
\Theta_{n}(x) & -(\Omega_{n+1}(x)+V(x)h_{n})
\end{array}\right)\left(\begin{array}{cc}
x-S_{n} & -R_{n}  \\
      1 & 0
\end{array}\right) \nn \\
&   & - \frac{1}{Wh_{n-1}}\left(\begin{array}{cc}
x-S_{n} & -R_{n}  \\
1 & 0
\end{array}\right)\left(\begin{array}{cc}
\Omega_{n}(x) & -\Theta_{n}(x)  \\
\Theta_{n-1}(x) & -(\Omega_{n}(x)+V(x)h_{n-1})
\end{array}\right) \nn \\
&&
\end{eqnarray}
we can identify two distinct relations
\bse\label{eq:cbOT}\begin{eqnarray}
(x-S_{n})\left(\frac{\Omega_{n+1}}{h_{n}}-\frac{\Omega_{n}}{h_{n-1}}\right) & = & R_{n+1}\frac{\Theta_{n+1}}{h_{n+1}}-R_{n}\frac{\Theta_{n-1}}{h_{n-1}}+W \label{eq:lgOT} \\
(x-S_{n})\frac{\Theta_{n}}{h_{n}} & = & \frac{\Omega_{n+1}}{h_{n}}+\frac{\Omega_{n}}{h_{n-1}}+V\label{eq:smOT}\ ,
\end{eqnarray}\ese
which we can identify as being comparable with the Laguerre-Freud equations \cite{BR}.

{\bf Remark 2.1}
We should point out that this system could be explored independent of orthogonal polynomials by simply setting
$V=v_{0}+v_{1}x+v_{2}x^2+\ldots+v_{n}x^n$ and $W=w_{0}+w_{1}x+w_{2}x^2+\ldots+w_{n}x^n$ (where the $v_{j},w_{j}$ are constants)
and then see what difference equations are produced for different orders of $V$ and $W$.
However since we are interested with the connections with semi-classical orthogonal polynomials, we will present a semi-classical weight
and then determine $V$ and $W$.


\section{A coupled difference equation and corresponding third order nonlinear equation}
\setcounter{equation}{0}

This method can be demonstrated by using a semi-classical weight $l_{0}(x)=(x-t)^{\ar}e^{-(ax+\frac{b}{2}x^{2})}$ 
synonymous with the (associated) Laguerre orthogonal polynomials $l(x)=x^{\ar}e^{-x}$.
Our choice of deformations for this weight, involve altering the order of the polynomial
in the exponential.

We first consider a deformation in the exponential part of the weight function,
the semi-classical weight $w(x)=(x-t)^{\ar}e^{-(ax+\frac{b}{2}x^{2})}$ with $\ar,a,b>0$
and where the support $S$ is an arc from $(t\rightarrow\infty)$.
Then from the Pearson equation, we have
\be
V(x) =  \ar-(a+bx)(x-t)\quad ,\quad W(x)  =  x-t.
\ee
and from the consistency relations we have two non-trivial equations
\bse\label{eq:Lagsf}\bea
&& b(R_{n+1}+R_{n})=-S_{n}[bS_{n}+(a-bt)]+(2n+1+at+\ar), \label{eq:Lagfent} \\
&& R_{n+1}[b(S_{n+1}+S_{n})+(a-bt)]-R_{n}[b(S_{n}+S_{n-1})+(a-bt)]=S_{n}-t. \nn  \\
&&\label{eq:Lagsent}
\eea\ese
We consider this to be a nonlinear system in terms of the recurrence coefficients $R_{n}$ and $S_{n}$,
which has the linear system (\ref{eq:LP}) with the Lax pair:
\bse\bea
L_{n}&=&\left(\begin{array}{cc}
x-S_{n} & -R_{n}  \\
      1 & 0
\end{array}\right)\quad , \quad  \\
M_{n}&=& \frac{1}{x-t}
\left(\begin{array}{cc}
n-bR_{n} & (bx+a+b(S_{n}-t))R_{n}  \\
-(bx+a+b(S_{n-1}-t)) & bx^2+x(a-bt)+bR_{n}-n-\ar-at
\end{array}\right) \nn  \\
&&
\eea\ese
for the associated semi-discrete Lax equation (\ref{eq:Lax}).  This system can be called a discrete integrable system due to the existence of
the corresponding linear problem, i.e., the Lax pair.  In the strictest sense we cannot call the equation Painlev\'e since it is third order however there
are other examples of third order nonlinear difference equations that are known to be integrable \cite{Conte} .

Writing this system in matrix form
\bea
&&\left(\begin{array}{cc}
b & b  \\
b(S_{n+1}+S_{n})+(a-bt) & -[b(S_{n}+S_{n-1})+(a-bt)]
\end{array}\right)\left(\begin{array}{c}
R_{n+1}  \\
R_{n}
\end{array}\right)\nn\\
&=&\left(\begin{array}{c}
-S_{n}(bS_{n}+(a-bt))+(2n+1+at+\ar) \\
S_{n}-t
\end{array}\right)
\eea
allows us to solve the system in terms of $R_{n+1}$ and $R_{n}$ and hence we can find a third order difference equation in $S_{n}$
\bea
&&\left\{(S_{n}+S_{n-1}+\left(\frac{a}{b}-t\right))(S_{n}(S_{n}+\left(\frac{a}{b}-t\right)))-\frac{1}{b}(2n+1+at+\ar)((S_{n}+S_{n-1})\right. \nn \\
&&+\left.\left(\frac{a}{b}-t\right))-(S_{n}-t)\right\}\times\left\{-2\left(S_{n+1}+\frac{a}{b}-t\right)-(S_{n}+2S_{n+1}+S_{n+2})\right\} \nn  \\
&=& \left\{(S_{n+2}+S_{n+1}+\left(\frac{a}{b}-t\right))(S_{n+1}(S_{n+1}+\left(\frac{a}{b}-t\right)))-\frac{1}{b}(2n+1+at+\ar)((S_{n+2}+S_{n+1})\right.\nn \\
&&+\left.\left(\frac{a}{b}-t\right))-(S_{n+1}-t)\right\}\times\left\{-2\left(S_{n}+\frac{a}{b}-t\right)-(S_{n-1}+2S_{n}+S_{n+1})\right\}\ .
\eea

Alternatively by letting $S_{n}=Q_{n}-Q_{n-1}$ in (\ref{eq:Lagsf}), we are led to an alternative third order difference equation in $Q_{n}$ that includes
an extra parameter $c$
\bea
&&b\left(\frac{Q_{n}-(n+1)t+c}{a-bt+b(Q_{n+1}-Q_{n-1})}+\frac{Q_{n-1}-nt+c}{a-bt+b(Q_{n}-Q_{n-2})}\right) \nn  \\
&=&-(Q_{n}-Q_{n-1})[b(Q_{n}-Q_{n-1})+a-bt]+(2n+1+at+\ar)\ ,
\eea
(where $c$ is the constant of integration).



\section{Conclusion and Outlook}

Given a class of semi-classical orthogonal polynomials (Hermite, Laguerre and Jacobi), we can identify a semi-discrete Lax pair and
thus an associated discrete integrable system.  Using the Laguerre weight $l_{0}(x)=(x-t)^{\ar}e^{-(ax+\frac{b}{2}x^{2})}$,
we found a new coupled discrete integrable system, which is first order in $R$ and second order in $S$.  Combining the two equations gives
a third order difference equation in $S$ or a third order difference equation in the new variable $Q$.
Since we were only interested in the connections with discrete Painlev\'e equations, we have omitted to look at the additional
$t$-differential equation (which appears as a consequence of the $t$ parameter in the weight function).  It is likely that
we could use the $t$-differential equation to find a continuous Painlev\'e equation related to our semi-classical Laguerre weight.

While we have chosen to look at a simple deformation of the classical orthogonal polynomial weight associated Laguerre, this scheme can of course be
used for deriving a multitude of discrete integrable systems, through choosing an appropriate classical weight function.  While we have made some
progress in this regard, finding the corresponding continuous equation is not always possible.
Working with further examples has shown that when looking for continuum limits, the choice of a semi-classical
weight function must be of a particular form.

In this paper we have applied the Laguerre method to a family of classical orthogonal polynomials, but we expect it is possible that the
Laguerre method can be used with other classes of orthogonal polynomials, such as the discrete, multiple or q-orthogonal polynomials.
We would need to alter the method appropriately, such as choosing an analogue for the Pearson equation (since we are using
the ``classical Pearson equation'' for classical orthogonal polynomials).  Thus, we could derive a similar scheme for
q-orthogonal polynomials given the q-Pearson equation, where a natural extension of this would be to consider the q-Laguerre orthogonal polynomials.

\section{Acknowledgements}

Paul Spicer wishes to thank Nalini Joshi for numerous discussions on integrable lattice
equations, and for advice and encouragement. He wishes to thank Pavlos Kassotakis for proof reading the paper and useful insights.
The author is supported by the Australian Research Council Discovery Project Grant \#DP0664624.

\end{document}